\documentstyle[epsfig,twoside,fleqn,ws2crc2]{article}

\newcommand{\simlt}  {\raisebox{-.6ex}{$\stackrel{\textstyle <}{\sim}$}}
\newcommand{\simgt}  {\raisebox{-.6ex}{$\stackrel{\textstyle >}{\sim}$}}
\newcommand{\AmS}{{\protect\the\textfont2
  A\kern-.1667em\lower.5ex\hbox{M}\kern-.125emS}}

\title{  {
         \small Topical Seminar
         on Neutrinos and Astroparticle Physics. 
         San Miniato. May 1999.}
         \hspace{0.8cm} 
         {\normalsize RAL-TR/1999 -066}\\
         \vspace{0.3cm} % \\
Tri-Maximal vs.\ Bi-Maximal
          Neutrino Mixing  % \hspace{2.5cm}hep-ph/9909431
}

\author{W. G. Scott \\ Rutherford Appleton Laboratory, 
        Chilton, Didcot, UK.}

\begin{document}

\begin{abstract}
It is argued that
data from atmospheric and solar neutrino experiments
point strongly to tri-maximal or bi-maximal lepton mixing.
While (`optimised') 
bi-maximal mixing gives an excellent
{\it a posteriori} fit to the data,
tri-maximal mixing
is an {\it a priori} hypothesis,
which is not excluded,
taking account of
terrestrial matter effects.
%are taken into account.
\end{abstract}

\maketitle
\vspace{-0.3cm}
\section{TRI-MAXIMAL MIXING}

Threefold maximal mixing, 
ie.\ tri-maximal mixing,
undeniably occupies a special place
in the space of 
all $3 \times 3$ mixings.
In some weak basis
the two non-commuting mass-matrices
$m^2_l$, $m^2_{\nu}$ 
($m_im_i^{\dagger} \equiv m^2_i$)
may simultaneously be written \cite{hs94}
as `circulative' matrices 
(`of degree zero') \cite{chen92}:
\begin{equation}
\left(\matrix{
a_l & b_l & \bar{b}_l \omega \cr
\bar{b}_l & a_l & b_l \omega \cr
b_l \bar{\omega} & \bar{b}_l \bar{\omega} & a_l \cr
} \right)
\hspace{0.7cm}
\left(\matrix{
a_{\nu} & b_{\nu} & \bar{b}_{\nu} \bar{\omega} \cr
\bar{b}_{\nu} & a_{\nu} & b_{\nu} \bar{\omega} \cr
b_{\nu} \omega & \bar{b}_{\nu} \omega & a_{\nu} \cr
} \right)
\end{equation}
respectively invariant under
monomial cyclic permutations 
(`circulation' matrices \cite{chen92})
of the form: 
\begin{equation}
\hspace{0.2cm} C=\left(\matrix{
. & 1 & . \cr
. & . & \omega \cr
\bar{\omega} & . & . \cr
} \right)
\hspace{0.4cm}
\bar{C}=\left(\matrix{
. & 1 & . \cr
. & . & \bar{\omega} \cr
\omega & . & . \cr
} \right)
\end{equation}
($C^{\dagger}m^2_lC = m^2_l$,  etc.)\
%$\bar{C}^{\dagger}m^2_{\nu}\bar{C} = m^2_{\nu}$)
just as circulant matrices are invariant 
under simple cyclic permutations \cite{adler}.
The mass matrices Eq.~1
are diagonalised by 
the (eg.\ circulant) unitary matrices
$V$ and $\bar{V}$: 
\begin{equation}
\frac{1}{\sqrt{3}} \left(\matrix{
1 & \bar{\omega} & 1 \cr
1 & 1 & \bar{\omega} \cr
\bar{\omega} & 1 & 1 \cr
} \right)
\hspace{0.9cm}
\frac{1}{\sqrt{3}} \left(\matrix{
1 & \omega & 1 \cr
1 & 1 & \omega \cr
\omega & 1 & 1 \cr
} \right)
\end{equation}
respectively,
leading to
threefold maximal mixing:
\begin{equation}
V \bar{V}^{\dagger} =
\frac{1}{3}\left(\matrix{
2+\omega & 
               2\bar{\omega}+1 & 
                              2\bar{\omega}+1 \cr
2\bar{\omega}+1 & 
               2+\omega & 
                              2\bar{\omega}+1 \cr
2\bar{\omega}+1 & 
               2\bar{\omega}+1 & 
                              2+\omega \cr
} \right)
\end{equation}
where in all the above
and in what follows 
$\omega$ and $\bar{\omega}$ 
represent complex cube-roots of unity
and the overhead 
`bar' denotes complex conjugation.

%With a more convenient choice of phases
After some rephasing
of rows and columns
the tri-maximal mixing matrix
may be re-written:
\begin{equation}
U \hspace{2mm} = \hspace{2mm} \frac{1}{\sqrt{3}} \hspace{1mm}
\matrix{ & \hspace{-0.4cm} \matrix{ \nu_1 & \nu_2 & \nu_3 } \cr
\matrix{ e \cr \mu \cr \tau} &
\hspace{-0.4cm} \left(\matrix{
1 & 
               1 & 
                              1 \cr
1 & 
               \omega & 
                              \bar{\omega} \cr
1 & 
               \bar{\omega} & 
                              \omega \cr
} \right) }
\end{equation}
where the matrix
in the parenthesis
is identically the character table
for the cyclic group $C_3$
(group elements vs.\ 
irreducible representations).
In threefold maximal mixing
the CP violation parameter
$|J_{CP}|$ is maximal
($|J_{CP}|=1/6\sqrt{3}$)
and if no two neutrinos
are degenerate in mass,
CP and T violating asymmetries
can approach $\pm 100\%$.

Observables depend
on the sqaures of the
moduli of the mixing matrix elements \cite{hps}:
\begin{equation}
(|U_{l\nu}|^2) \hspace{2mm} = \hspace{2mm}
%U \circ \bar{U} \hspace{0.2cm} = \hspace{0.2cm}
\matrix{    & \hspace{-0.4cm} \matrix{\nu_1 & \nu_2 & \nu_3} \cr
\matrix{  e \cr \mu \cr \tau \cr} & 
\hspace{-0.4cm} \left(\matrix{
1/3 & 
               1/3 & 
                              1/3 \cr
1/3 & 
               1/3 & 
                              1/3 \cr
1/3 & 
               1/3 & 
                              1/3 \cr
} \right) }.
\end{equation}
%(where $\circ$ denotes the entrywise matrix product).
In tri-maximal mixing
all survival and appearance probabilities
are universal (ie.\ flavour independent)
and in particular if two neutrinos
are effectively degenerate
tri-maximal mixing predicts
for the locally averaged 
survival probability:
\begin{equation}
<\hspace{-1mm} P(l \rightarrow l) \hspace{-1mm}> 
%\hspace{1mm} = \hspace{1mm}
\hspace{1mm} = 
(1/3+1/3)^2 +(1/3)^2 
%\hspace{1mm} = \hspace{1mm}
= 
5/9 % \hspace{1.0cm} (l = l)
\end{equation}
and appearnce probability: %($l \ne l'$):
\begin{equation}
<\hspace{-1mm} P(l \rightarrow l') \hspace{-1mm}> 
\hspace{1mm} = % \hspace{1mm} 2/9 \hspace{1.0cm} (l \ne l')
2 \times (1/3)(1/3) =2/9.
\end{equation}
If all three neutrino masses are effectively non-degenerate:
$< \hspace{-1mm} P(l \rightarrow l) \hspace{-1mm}> 
\hspace{1mm} = \hspace{1mm} 
< \hspace{-1mm} P( l \rightarrow l') \hspace{-1mm} > 
\hspace{1mm} = \hspace{1mm}1/3$.
\vspace{-0.5cm}
\section{ATMOSPHERIC NEUTRINOS}

The SUPER-K \cite{walt} multi-GeV data,
show a clear $\sim50$\% suppression of atmospheric $\nu_{\mu}$
for zenith angles $\cos \theta$ $\simlt$ 0,
as shown in Fig.~1a.
At the same time 
the corresponding distribution for $\nu_e$
seems to be very largely unaffected,
as shown in Fig.~1b.
The best fit
is for (twofold) maximal $\nu_{\mu}-\nu_{\tau}$ mixing
with $\Delta m^2 \simeq 3.0 \times 10^{-3}$ eV$^2$,
as shown by the solid/dotted curves in Fig.~1.
\vspace{-0.4cm}
\begin{figure}[h]
\epsfig{figure=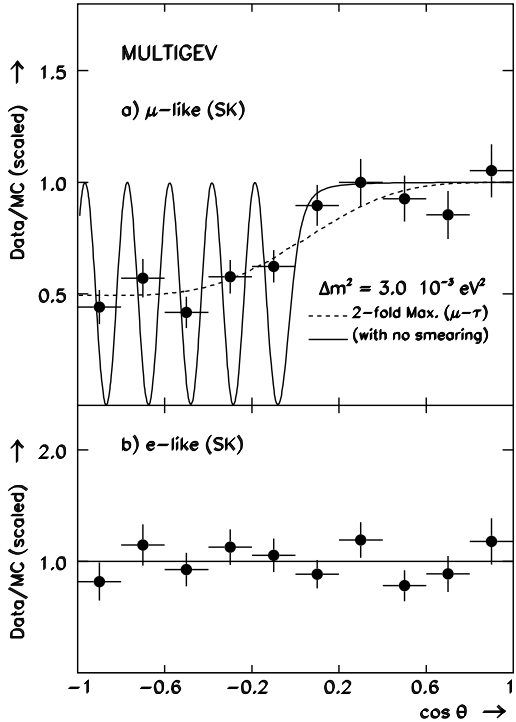,width=85mm,bbllx=100pt,bblly=210pt
,bburx=550pt,bbury=690pt}
\caption{The multi-GeV zenith-angle distributions
for a) $\mu$-like and b) $e$-like events
from the SUPER-K experiment.
The solid curve is for
maximal $\nu_{\mu}-\nu_{\tau}$ oscillations
with $\Delta m^2  = 3.0 \times 10^{-3}$ eV$^2$
and the dotted curve shows the effect
of energy averaging and angular smearing.}
\end{figure}
\vspace{-0.6cm}

\noindent
From the measured suppression we have: 
\begin{equation}
(1-|U_{\mu 3}|^2)^2+(|U_{\mu 3}|^2)^2 \simeq 0.52 \pm 0.05
\end{equation}
whereby
the $\nu_{\mu}$
must have a large $\nu_3$ content,
ie.\ $|U_{\mu 3}|^2 \simeq 1/2$, or more precisely:
\begin{equation}
1/3 \hspace{3mm} \simlt \hspace{3mm} |U_{\mu 3}|^2 
\hspace{3mm} \simlt \hspace{3mm} 2/3
\end{equation}
where the range quoted corresponds
to the $1\sigma$ error above
(which is largely statistical).

\section{THE CHOOZ DATA}

The apparent lack
of $\nu_e$ mixing
at the atmospheric scale,
is supported by independent data
from the CHOOZ reactor \cite{chz99} (Fig.~2),
which rules out large $\nu_e$ mixing
over most of the $\Delta m^2$ range
allowed in the 
atmospheric neutrino experiments.

\vspace{-0.6cm}
\begin{figure}[h]
\epsfig{figure=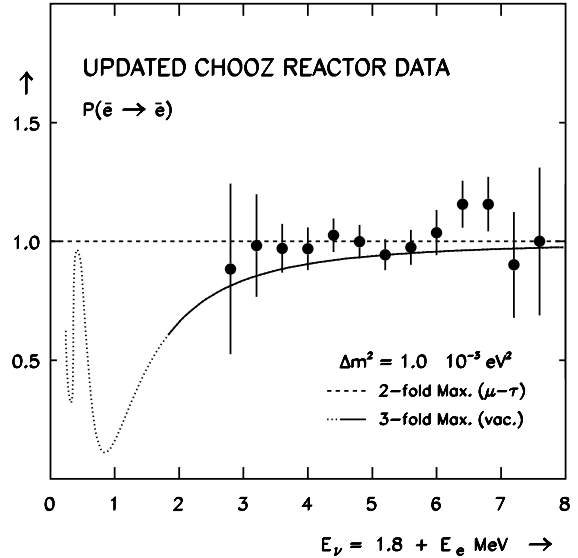,width=85mm,bbllx=100pt,bblly=270pt
,bburx=550pt,bbury=630pt}
\caption{The latest data 
from the CHOOZ reactor,
corresponding to the full data taking.
(The solid curve is
for tri-maximal mixing
with $\Delta m^2  = 1.0 \times 10^{-3}$ eV$^2$.)}
\end{figure}
\vspace{-0.6cm}

\noindent Taken together,
the CHOOZ and SUPER-K data indicate
that the $\nu_3$ has no $\nu_e$ content,
ie.\ there is a zero (or near-zero)
in the top right-hand corner of the
lepton mixing matrix,
$|U_{e 3}|^2$ $\simlt$ $0.03$.

\subsection{The Fritzsch-Xing Ansatz}

Remarkably,
the Fritzsch-Xing hypothesis \cite{xing}
(published well before
the initial CHOOZ data) 
predicted just such a zero:
\begin{equation}
(|U_{l\nu}|^2) \hspace{2mm} = \hspace{2mm}
%U \circ \bar{U} \hspace{0.2cm} = \hspace{0.2cm}
\matrix{    & \hspace{-0.4cm} \matrix{\nu_1 & \nu_2 & \nu_3} \cr
\matrix{  e \cr \mu \cr \tau \cr} & 
\hspace{-0.4cm} \left(\matrix{
1/2 & 
               1/2 & 
                              . \cr
1/6 & 
               1/6 & 
                              2/3 \cr
1/3 & 
               1/3 & 
                              1/3 \cr
} \right) }
\end{equation}
The Fritzsch-Xing ansatz
(Eq.~11) is readily obtained from
theeefold maximal mixing (Eq.~5)
by the re-definitions:
$\nu_e \rightarrow (\nu_e-\nu_{\mu})/\sqrt{2}$
and
$\nu_{\mu} \rightarrow (\nu_e+\nu_{\mu})/\sqrt{2}$
(up to phases),
keeping the $\nu_{\tau}$ tri-maximally mixed.
While the {\it a priori} argument
for these particular linear combinations \cite{xing}
is far from convincing,
it is clear that Eq.~11 is so far consistent
with the atmospheric data:
\begin{equation}
<\hspace{-1mm} P(\mu \rightarrow \mu) \hspace{-1mm}> 
%\hspace{1mm} = \hspace{1mm}
\hspace{1mm} = 
(1/6+1/6)^2 +(2/3)^2 
%\hspace{1mm} = \hspace{1mm}
= 
5/9 % \hspace{1.0cm} (l = l)
\end{equation}
(cf.\ Eq.~9), while beyond the second threshold:
\begin{equation}
<\hspace{-1mm} P(e \rightarrow e) \hspace{-1mm}> 
%\hspace{1mm} = \hspace{1mm}
\hspace{1mm} = 
(1/2)^2+(1/2)^2 +(0)^2 
%\hspace{1mm} = \hspace{1mm}
= 1/2
\end{equation}
The famous `bi-maximal' scheme \cite{gold}
is very similar to the Fritzsch-Xing ansatz
and likewise predicts a $50$\% suppression
for the solar data.

\section{THE SOLAR DATA}

Taken at face value,
the results from the various solar neutrino experiments
imply an energy dependent suppression.
In particular, taking BP98 fluxes
(and correcting for the NC contribution in SUPER-K), 
the results from HOMESTAKE and SUPER-K:
$P(e \rightarrow e) \simeq 0.3-0.4$,
lie significantly below the results 
from the gallium experiments:
$P(e \rightarrow e) \simeq 0.5-0.6$,
as shown in Fig.~3.

\subsection{`Optimised' Bi-Maximal Mixing}

Mindful of the
popularity of the large-angle MSW solution 
and the undenied phenomenological promise
of the `original' bi-maximal scheme \cite{gold},
we have ourselves proposed \cite{hps},
a phenomenologically viable 
(and even phenomenologically favoured)
{\it a posteriori} \hspace{0.5mm}
`straw-man' alternative to tri-maximal mixing:

\begin{equation}
(|U_{l\nu}|^2) \hspace{2mm} = \hspace{2mm}
%U \circ \bar{U} \hspace{0.2cm} = \hspace{0.2cm}
\matrix{    & \hspace{-0.4cm} \matrix{\nu_1 & \nu_2 & \nu_3} \cr
\matrix{  e \cr \mu \cr \tau \cr} & 
\hspace{-0.4cm} \left(\matrix{
2/3 & 
               1/3 & 
                              . \cr
1/6 & 
               1/3 & 
                              1/2 \cr
1/6 & 
               1/3 & 
                              1/2 \cr
} \right) }
\end{equation}

\noindent which we refer to here as `optimised' bi-maximal mixing.
This scheme is of course just one
special case of the `generalised'
bi-naximal hypotheses 
of Altarelli and Feruglio \cite{alt}
(and see also Ref. \cite{jarl}).
At the atmospheric scale Eq.~14 gives:
\begin{equation}
<\hspace{-1mm} P(\mu \rightarrow \mu) \hspace{-1mm}> 
%\hspace{1mm} = \hspace{1mm}
\hspace{1mm} = 
(1/6+1/3)^2 +(1/2)^2 
%\hspace{1mm} = \hspace{1mm}
= 1/2
\end{equation}
while beyond the second scale it gives:
\begin{equation}
<\hspace{-1mm} P(e \rightarrow e) \hspace{-1mm}> 
%\hspace{1mm} = \hspace{1mm}
\hspace{1mm} = 
(2/3)^2+(1/3)^2 +(0)^2 
%\hspace{1mm} = \hspace{1mm}
= 5/9
\end{equation}
There is then
the added possibility to exploit
a large angle MSW solution
with the base of the `bathtub'
(given by the $\nu_e$ content of the $\nu_2$)
given by $P(e \rightarrow e) = 1/3$,
as shown in Fig.~3.
\vspace{-.4cm}
\begin{figure}[h]
\epsfig{figure=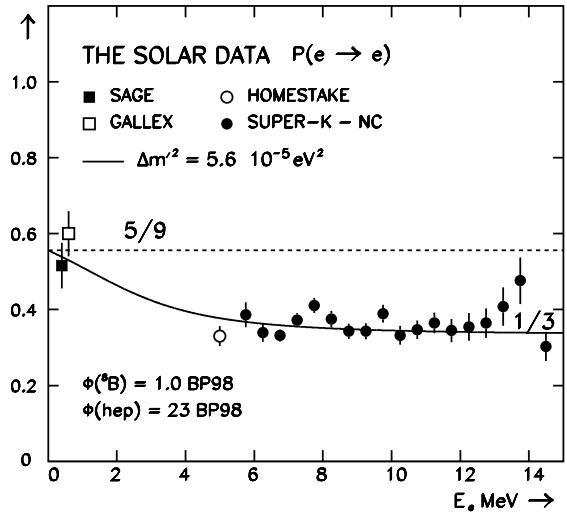,width=85mm,bbllx=100pt,bblly=260pt
,bburx=550pt,bbury=600pt}
\caption{The SUPER-K solar data [11] 
plotted as a function of recoil electron energy $E_e$.
The results from the two gallium experiments 
SAGE and GALLEX and
the HOMESTAKE experiment are also plotted 
(at $< \hspace{-1.5mm} 1/E \hspace{-1.5mm} >^{-1} \sim 0.5$ MeV
and $< \hspace{-1.5mm} 1/E \hspace{-1.5mm} >^{-1} \sim 5$ MeV respectively).
%All points plotted using BP98 fluxes [12].
Used BP98 [12] fluxes (with rescaled hep).
The SUPER-K points are corrected
for the NC contribution.
The solid curve is Eq.~14
with $\Delta m'^2 = 5.6 \times 10^{-5}$ eV$^2$.}
\end{figure}

\vspace{-0.6cm}
Although clearly the matrix Eq.~14
is readily obtained from the
tri-maximal mixing matrix Eq.~5
by forming linear combinations
of the heaviest and lightest
mass eigenstates:
$\nu_1 \rightarrow (\nu_1 + \nu_3)/\sqrt{2}$ and
$\nu_3 \rightarrow (\nu_1 - \nu_3)/\sqrt{2}$
(up to phases),
with the $\nu_2$ remaining tri-maximally mixed,
we emphasise that
we claim no understanding
of why these redefinitions  should be necessary.
%and have arrived at this
%proposal in a purely ad hoc manner.

\section{TERRESTRIAL MATTER EFFECTS \\ 
         IN TRI-MAXIMAL MIXING}

%In the tri-maximal scenario, 
%mattter effects in the Sun 
%are expected to have
%little observable influence.
%On the other hand,
%terrestrial matter effects
%for atmospheric neutrinos
%can be very importnat
%in tri-maximal mixing.

In general,
matter effects deform
the mixing matrix
and shift the neutrino masses,
away from their vacuum values,
depending on the local matter density.
In the tri-maximal mixing scenario,
the CHOOZ data require 
$\Delta m^2$ $\simlt$ $10^{-3}$ eV$^2$ (Fig~2),
so that matter effects
can be very important. 
For `intermeduate' densities \cite{hps},
%matter effects remove the effective degeneracy
%between the light neutrinos
\vspace{-0.4cm}
\begin{figure}[h]
\epsfig{figure=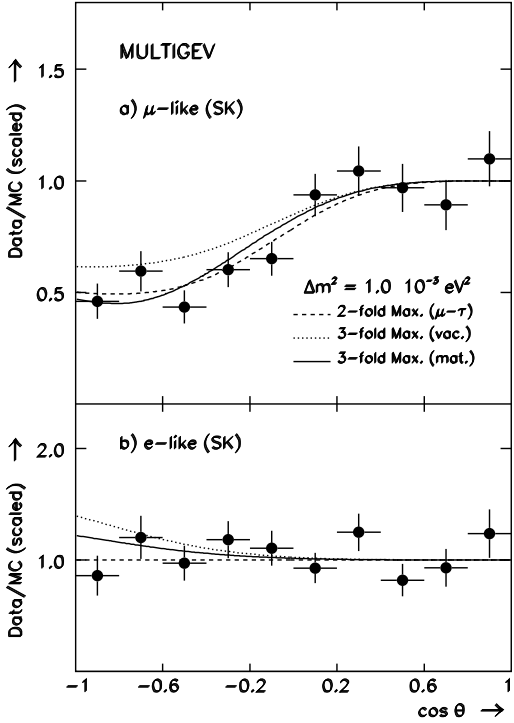,width=85mm,bbllx=100pt,bblly=210pt
,bburx=550pt,bbury=690pt}
\caption{The multi-GeV zenith-angle distributions
for a) $\mu$-like and b) $e$-like events
from the SUPER-K experiment.
Tri-maximal mixing with matter effects
(solid curve) 
is closer to twofold
maximal $\nu_{\mu}-\nu_{\tau}$ mixing
(dashed curve)
than to vacuum tri-maximal mixing 
(dotted curve).}
\end{figure}
\vspace{-0.7cm}
%such that 
the matter mass eigenstates become:
$\nu_1 \rightarrow (\nu_1 - \nu_2)/\sqrt{2}$ and
$\nu_2 \rightarrow (\nu_1 + \nu_2)/\sqrt{2}$
(up to phases)
while the $\nu_3$ remains tri-maximally mixed:
\begin{equation}
(|U_{l\nu}|^2) \hspace{2mm} \rightarrow \hspace{2mm}
%U \circ \bar{U} \hspace{0.2cm} = \hspace{0.2cm}
\matrix{    & \hspace{-0.4cm} \matrix{\nu_1 & \nu_2 & \nu_3} \cr
\matrix{  e \cr \mu \cr \tau \cr} & 
\hspace{-0.4cm} \left(\matrix{
. & 
               2/3 & 
                              1/3 \cr
1/2 & 
               1/6 & 
                              1/3 \cr
1/2 & 
               1/6 & 
                              1/3 \cr
} \right) }
\end{equation}

As evidenced in Fig.~4, 
the phenomenology of Eq.~17
can be almost indistinguishable
from that of 
`optimised' bi-maximal mixing, Eq.~14.
Beyond the `matter threshold'
we have for $\nu_{\mu}$: 
\begin{equation}
<\hspace{-1.3mm} 
    P(\mu \hspace{-0.5mm} \rightarrow \hspace{-0.5mm} \mu) 
                                              \hspace{-1.4mm}> 
\hspace{0.1mm} = \hspace{-1mm}
(1/2)^2+(1/6)^2 +(1/3)^2 
= \hspace{-1mm} 7/18
\end{equation}
while for $\nu_e$:
\begin{equation}
<\hspace{-1mm} P(e \rightarrow e) \hspace{-1mm}> 
\hspace{1mm} = \hspace{-1mm}
(0)^2+(2/3)^2 +(1/3)^2 
= 5/9
\end{equation}

For atmospheric neutrinos,
account must be taken of
the initial flux ratio.
$\phi(\nu_{\mu})/\phi(\nu_e)$.
For $\phi(\nu_{\mu})/\phi(\nu_e)$ $\simeq$ $2/1$,
the effective $\nu_{\mu}$ suppression:
\begin{equation}
7/18 + 1/2 \times 2/9 = 1/2
\end{equation}
(cf.\ Eq.~15) while the $\nu_e$ rate
is fully compensated:
\begin{equation}
5/9 +1/2 \times 2/9 = 1
\end{equation}
so that $\nu_e$ appear not
to participate in the mixing.
\vspace{-0.4cm}
\begin{figure}[h]
\epsfig{figure=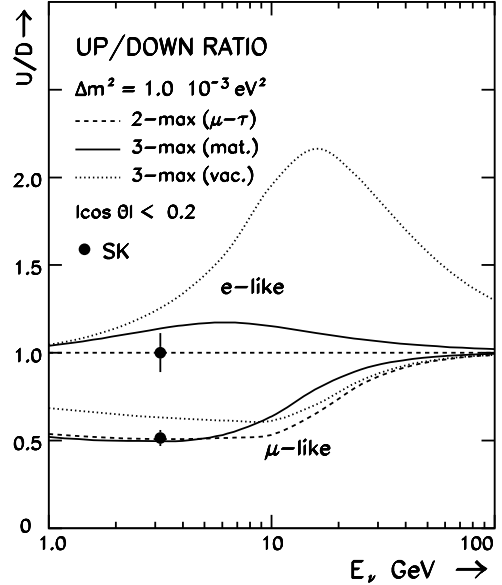,width=85mm,bbllx=100pt,bblly=270pt
,bburx=540pt,bbury=600pt}
\caption{The up/down ratio
for multi-GeV $e$-like and $\mu$-like events
as measured by SUPER-K.
The curves are the various expectations
plotted vs.\ neutrino energy,
for $\Delta m^2  = 1.00 \times 10^{-3}$ eV$^2$.}
\end{figure}
\vspace{-0.5cm}

The up/down ratio (Fig.~5)
measures the effective supression.
For energies $E$ $\simgt$ $1$ GeV
the initial flux ratio
$\phi(\nu_{\mu})/\phi(\nu_e)$ $\simgt$ $2/1$
and the $\nu_e$ rate
becomes `over-compensated',
while, for sufficiently
high energies
compensation effects vanish
as the  complete decoupling limit
$\nu_e \rightarrow \nu_3$
is approached.
The resulting maximum
in the up/down ratio for $\nu_e$
(Fig.~5)
is describerd as a `resonance'
by Pantaleone \cite{pant}.

\subsection{Matter Induced CP-violation}

As regards the mixing matrix,
CP and T violating effects are maximal
in tri-maximal mixing,
but in spite of this,
due to the extreme hierarchy
of $\Delta m^2$ values involved,
for most accessible $L/E$ values,
observable asymmetries 
are expected to be unmeasurably small
(in vacuum).

\vspace{-1.0cm}
\begin{figure}[h]
\epsfig{figure=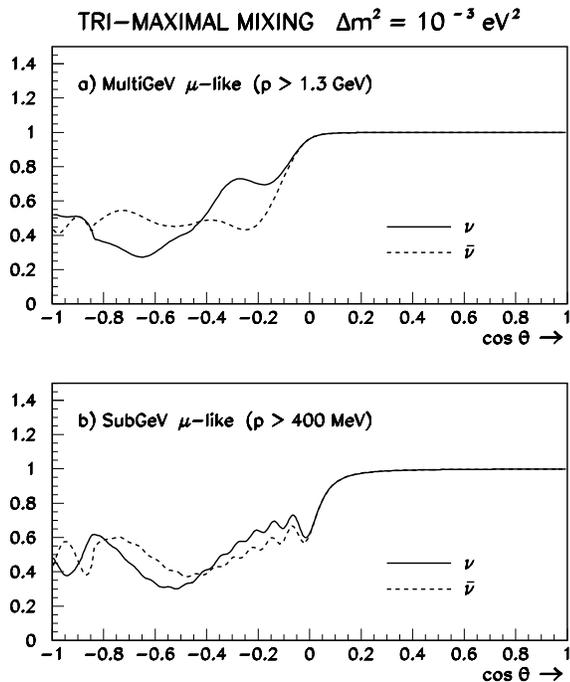,width=85mm,bbllx=100pt,bblly=210pt
,bburx=550pt,bbury=690pt}
\caption{Predicted zenith-angle distributions
in tri-maximal mixing,
for a) multi-GeV and b) sub-GeV events
in an atmopheric neutrinoexperiment,
seperating $\nu$ (solid curve) and $\bar{\nu}$ (dashed curve)
contributions.
The curves plotted include
energy averaging, but {\em not} angular smearing.}
\end{figure}

\vspace{-0.6cm}
In the presence of matter
(or indeed anti-matter)
significant asymmetries can occur,
however,
`enhanced' or
`induced' by matter effects.
Thus for example 
if atmospheric neutrinos
are separated $\nu/\bar{\nu}$,
interesting 
matter induced asymmetries
become observable (Fig.~6)
in tri-maximal mixing.
Such effects could be investigated
using atmospheric neutrino detectors 
equipped with magnetic fields \cite{mono},
and/or by using sign-selected beams
in long-baseline experiments \cite{min}.

%\vspace{-0.4cm}
\section{TRI-MAXIMAL MIXING AND \\
         THE SOLAR DATA}

The vacuum predictions
for the solar data
in tri-maximal mixing
are largely unmodified by matter effects
in the Sun, as is well known,
or, as we have seen, 
by matter effects 
in the Earth (Eq.~7 vs.\ Eq.~16).
Thus we expect
$P(e \rightarrow e) =5/9$ 
in tri-maximal mixing
with no energy dependence,
as shown in Fig.~7.
(Note that also the
\vspace{-0.6cm}
\begin{figure}[h]
\epsfig{figure=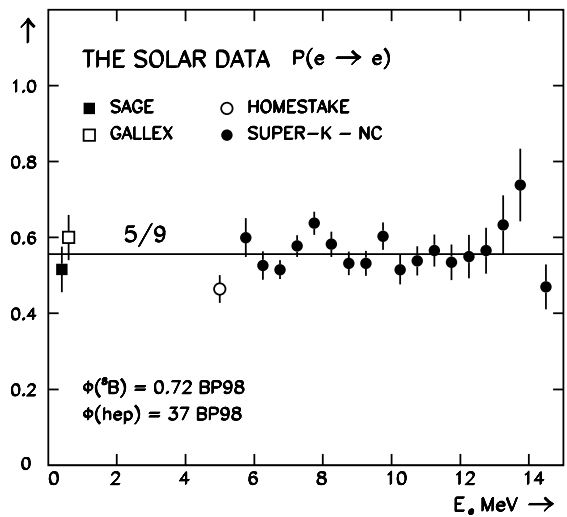,width=85mm,bbllx=100pt,bblly=260pt
,bburx=550pt,bbury=600pt}
\caption{The SUPER-K solar data [11]
plotted as a function of recoil electron energy $E_e$.
The results from the two gallium experiments 
SAGE and GALLEX and
the HOMESTAKE experiment are also plotted 
(at $< \hspace{-1.5mm} 1/E \hspace{-1.5mm} >^{-1} \sim 0.5$ MeV
and $< \hspace{-1.5mm} 1/E \hspace{-1.5mm} >^{-1} \sim 5$ MeV respectively).
Note that the SUPER-K points have been
corrected for the NC contribution, and that 
the non-pp fluxes have been freely rescaled,
with respect to BP98 [12],
to test for an energy independent suppression.}
\end{figure}
\vspace{-0.5cm}
`optimsed' bi-maximal scheme
%Eq.~14, 
predicts
$P(e \rightarrow e) =5/9$
with no energy dependence
for $\Delta m'^2$ 
outside the `bathtub').
In Fig.~7, the
$^8$B (and $^7$Be) flux,
affecting both the SUPER-K and HOMESTAKE data-points,
has been rescaled by an arbitrary
factor (0.72) for comparison
to the energy-independent prediction.
Given the flux errors 
($\sim \pm 14\%$ for $^8$B \cite{bp98}),
the fit (Fig.~7) seems not unreasonable.

\begin{figure*}[ht]
\epsfig{figure=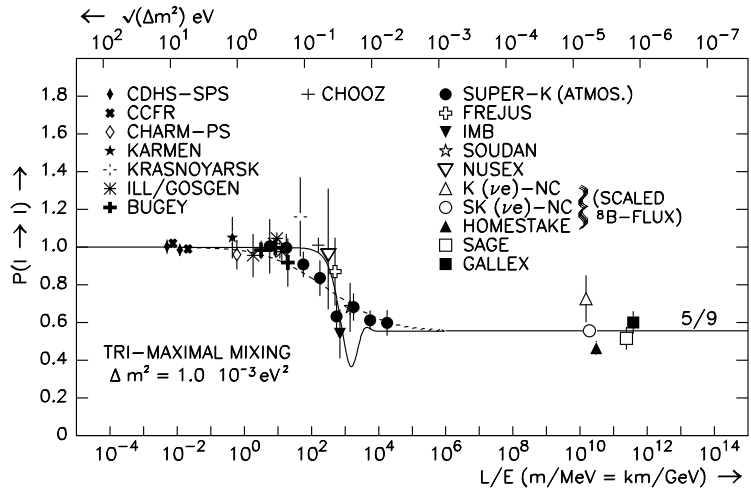,width=110mm,bbllx=-100pt,bblly=320pt
,bburx=450pt,bbury=600pt}
\caption{The updated $L/E$ plot for disappearance experiments.
The solid curve represents tri-maximal mixing with 
$\Delta m^2 = 1.0 \times 10^{-3}$ eV$^2$ and the dashed curve indicates
the angular smearing in SUPER-K.}
\end{figure*}
\vspace{-0.3cm}
%It is
%perhaps then useful to update  
%the $L/E$ plot \cite{hps},
%showing the
Fig.~8 shows the
solar, atmospheric,
accelerator and reactor data
in overall perspective,
within the tri-maximal context.
Note that {\em dis}-appearance results
{\em only} \hspace{0.4mm} are represented:
%on the plot:
if the LSND appearance
result \cite{lsnd} 
were ever to be confirmed,
tri-maximal mixing
would be instantly excluded.

\vspace{-0.4cm}
\section{CONCLUSION}

\noindent The lepton mixing
really does look to be either
tri-maximal or bi-naximal
at this point.
Tri-maximal mixing
is currently `squeezed'
in $\Delta m^2$
by CHOOZ vs.\ SUPER-K (Fig.~1-2).
For some $\Delta m'^2$,
bi-maximal mixing
(in particular the
`optimised' version discussed here)
clearly fits the data better.
Tri-maximal mixing
remains, however,  the `simplest'
most `symmetric' 
possibility. 
 
\vspace{3mm}
\noindent {\bf Acknowledgement}

\vspace{2mm}
\noindent I wish to thank
Paul Harrison and Don Perkins
for continued collaboration,
and numerous helpful discussions 
relating to the above material.

\end{document}